\documentclass[%
 reprint,
 amsmath,amssymb,
 prl,
floatfix,
]{revtex4-1}

\usepackage{graphicx}
\usepackage{dcolumn}
\usepackage{bm}
\usepackage{ifthen}
\usepackage{multirow}

\usepackage[separate-uncertainty=true]{siunitx}
\DeclareSIUnit\gauss{G}

\newcommand{\tauGR}{\tau_{\rm GR}}
\newcommand{\tauLPI}{\tau_{\rm LPI}}
\newcommand{\tauDLF}{\tau_{\rm DLF}}

\newcommand{\tauData}{\tau_{\rm ESOC}}

\newcommand{\tauCorr}{\tau_{\rm corr}}

\begin{document}


\title{A gravitational redshift test using eccentric Galileo satellites}

\author{P. Delva$^1$}
\email{Pacome.Delva@obspm.fr}
\author{N. Puchades$^{2,1}$}
\author{E. Sch\"onemann$^3$}
\author{F. Dilssner$^3$}
\author{C. Courde$^4$}
\author{S. Bertone$^5$}
\author{F. Gonzalez$^6$}
\author{A. Hees$^1$}
\author{Ch. Le Poncin-Lafitte$^{1}$}
\author{F. Meynadier$^{1}$}
\author{R. Prieto-Cerdeira$^6$}
\author{B. Sohet$^{1}$}
\author{J. Ventura-Traveset$^7$}
\author{P. Wolf$^{1}$}

\affiliation{$^1$SYRTE, Observatoire de Paris, Universit\'e PSL, CNRS, Sorbonne Universit\'e, LNE, 61 avenue de l'Observatoire 75014 Paris France}
\affiliation{$^2$Departamento de Astronomía y Astrofísica, Edificio de Investigación Jerónimo Muñoz, C/ Dr. Moliner, 50, 46100 Burjassot (Valencia), España}
\affiliation{$^3$European Space Operations Center, ESA/ESOC, 64293 Darmstadt Germany}
\affiliation{$^4$UMR Geoazur, Universit\'e de Nice, Observatoire de la C\^ote d'Azur, 250 rue A. Einstein, F-06560 Valbonne, France}
\affiliation{$^5$Astronomical Institute, University of Bern, Sidlerstrasse 5 CH-3012 Bern, Switzerland}
\affiliation{$^6$European Space and Technology Centre, ESA/ESTEC, 2200 AG Noordwijk, The Netherlands}
\affiliation{$^7$European Space and Astronomy Center, ESA/ESAC, 28692 Villanueva de la Ca\~nada, Spain}
\date{\today}

\begin{abstract}
We report on a new test of the gravitational redshift and thus of local position invariance, an integral part of the Einstein equivalence principle, which is the foundation of general relativity and all metric theories of gravitation. We use data spanning 1008 days from two satellites of Galileo, Europe's global satellite navigation system (GNSS), which were launched in 2014, but accidentally delivered on elliptic rather than circular orbits. The resulting modulation of the gravitational redshift of the onboard atomic clocks allows the redshift determination with high accuracy. Additionally specific laser ranging campaigns to the two satellites have enabled a good estimation of systematic effects related to orbit uncertainties. Together with a careful conservative modelling and control of other systematic effects we measure the fractional deviation of the gravitational redshift from the prediction by general relativity to be $\num{+0.19 \pm 2.48 e-5}$ at 1 sigma, improving the best previous test by a factor~5.6. To our knowledge, this represents the first reported improvement on one of the longest standing results in experimental gravitation, the Gravity Probe A hydrogen maser rocket experiment back in 1976.

\end{abstract}

\pacs{Valid PACS appear here}
\maketitle

The classical theory of general relativity~(GR) provides a geometrical description of the gravitational interaction. It is based on two fundamental principles: (i) the Einstein equivalence principle~(EEP) and (ii) the Einstein field equations that can be derived from the Einstein-Hilbert action. Although very successful so far, there are reasons to think that sufficiently sensitive measurements could uncover a failure of GR. For example, the unification of gravitation with the other fundamental interactions, and quantum theories of gravitation, generally lead to small deviations from GR~(see e.g.~\cite{Will2014}). Also dark matter and energy are so far only observed through their gravitational effects, but might be hints towards a modification of GR~\cite{Famaey2012,Clifton2012}.

From a phenomenological point of view, three aspects of the EEP can be tested: (i) the universality of free fall~(UFF); (ii) local Lorentz invariance~(LLI); and (iii) local position invariance~(LPI). Constraints on UFF have been recently improved by the Microscope space mission~\cite{Touboul2017a}, while LLI was recently constrained, for example, by using a ground fibre network of optical clocks~\cite{Delva2017e} (see e.g.~\cite{Will2014,Mattingly2005,Safronova2017} for reviews). In this paper we focus on testing LPI.

LPI stipulates that the outcome of any local non-gravitational experiment is independent of the space-time position of the freely-falling reference frame in which it is performed. This principle is mainly tested by two types of experiments: (i) search for variations in the constants of Nature (see e.g.~\cite{Guena2012}, and~\cite{Uzan2011} for a review) and (ii) gravitational redshift tests. The gravitational redshift was observed in a ground experiment for the first time by Pound, Rebka and Snider~\cite{Pound1959,Pound1965}.

In a typical clock redshift experiment, the fractional frequency difference $z=\Delta\nu / \nu$ between two clocks located at different positions in a static gravitational field is measured, by exchange of electromagnetic signals. The EEP predicts $z = \Delta U / c^2$ for stationary clocks, where $\Delta U$ is the gravitational potential difference between the locations of both clocks, and $c$ is the velocity of light in vacuum. A simple and convenient formalism to test the gravitational redshift is to introduce a new parameter $\alpha$ defined through~(see e.g.~\cite{Will2014}):
\begin{equation}
	z = \frac{\Delta\nu}{\nu} = (1+\alpha) \frac{\Delta U}{c^2}
	\label{eq:redshift}
\end{equation}
with $\alpha$ vanishing when the EEP is valid.

So far, the most accurate test of the gravitational redshift has been realized with the Vessot-Levine rocket experiment in 1976, also named the Gravity Probe A (GP-A) experiment~\cite{Vessot1980,Vessot1979,Vessot1989}. The frequency differences between a space-borne hydrogen maser clock and ground hydrogen masers were measured thanks to a continuous two-way microwave link. The total duration of the experiment was limited to 2 hours constrained to the parabolic trajectory of the GP-A rocket, and reached an uncertainty of $|\alpha|\leq 1.4\times 10^{-4}$~\cite{Vessot1989}. The future Atomic Clock Ensemble in Space (ACES) experiment~\cite{Cacciapuoti2009a,Meynadier2018}, an ESA/CNES mission, planned to fly on the ISS in 2020, will test the gravitational redshift to around $|\alpha|\leq 3\times 10^{-6}$. Furthermore, other projects like STE-QUEST propose to test the gravitational redshift at the level of $10^{-7}$~\cite{Altschul2015a}. Finally, observations with the RadioAstron telescope are hoping to reach an uncertainty of the order of $10^{-5}$~\cite{Litvinov2017a}.

In this article, following the proposal in~\cite{Delva2015q}, we use the onboard atomic clocks of the Galileo satellites 5 and 6 (named Doresa and Milena, or GSAT0201 and GSAT0202) to search for violations of the EEP/LPI. These two satellites were launched together on a Soyuz Rocket on August, 22$^{\rm nd}$ 2014 and because of a technical problem on the launcher's upper stage, they were placed in a non-nominal elliptic orbit. Although the satellites' orbits were adjusted after the launch, they remain elliptical, with each satellite climbing and falling some \SI{8500}{\kilo\meter} twice per day. The elliptic orbit induces a periodic modulation of the gravitational redshift at orbital period (around 13~hours), while the good stability of recent GNSS clocks allows us to test this periodic modulation to a new level of uncertainty. The Galileo 5 and~6 satellites, with their large eccentricity ($e=0.162$) and onboard passive hydrogen-maser (PHM) clocks, are hence perfect candidates to perform this test. Contrary to the GP-A experiment, it is possible to integrate the signal over a long duration, therefore improving the statistics. Moreover, Satellite Laser Ranging (SLR) data are used for a characterization of systematic effects. A specific ILRS (International Laser Ranging Service) campaign took place during the years 2016--2017~\cite{Delva2016c}.

\begin{figure}[t]
	\includegraphics[width=\linewidth]{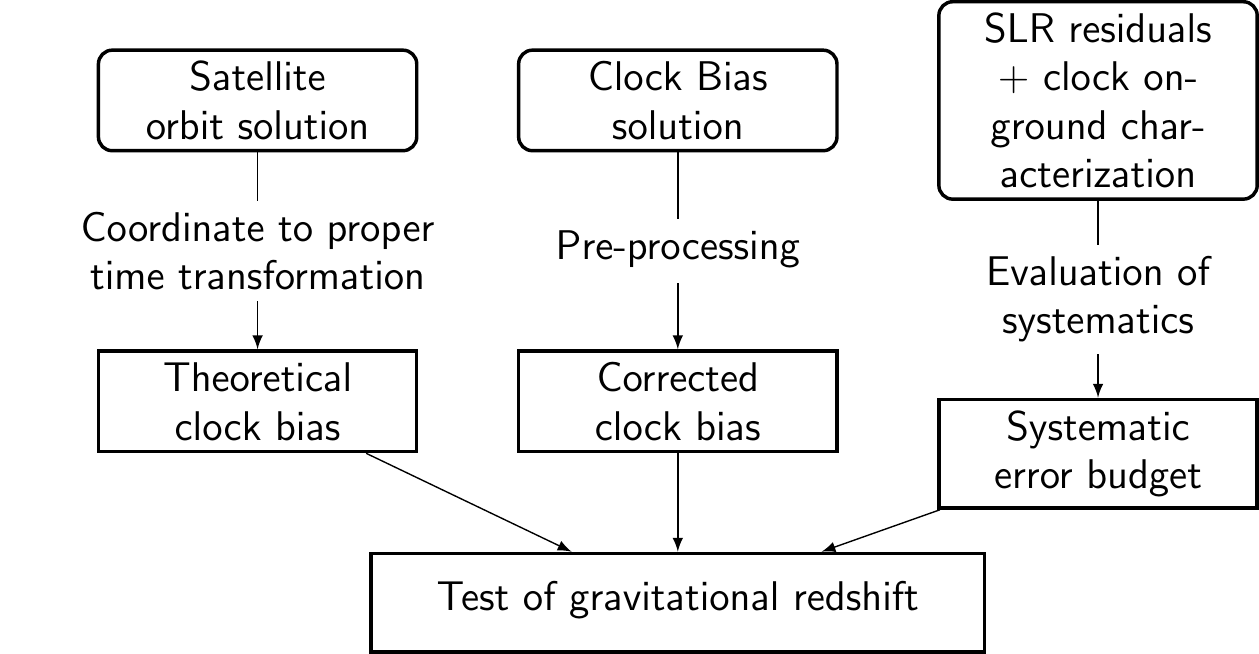}
	\caption{Data analysis flowchart: as input we use ESOC orbit and clock solution files, SLR residuals as well as clock on-ground characterization. The evaluation of systematics is completely independent from the clock measurements.}
	\label{fig:flowchart}
\end{figure}

The flowchart of the data analysis is given in Figure~\ref{fig:flowchart}. We use an orbit and clock solution generated by ESA's Navigation Support Office, located at the European Space Operations Centre (ESOC). The details of the ESOC processing strategy are given in the supplemental material~\cite{supp_mat}. The satellite orbit solution contains positions and velocities of multiple GNSS satellites in the terrestrial reference frame ITRF2014 with respect to GPS time epochs. Orbit solutions are independent of a possible violation of the gravitational redshift (at the required accuracy), as no assumptions or models of the clock evolution, are made. Instead the clock solutions are obtained as a free parameter for each epoch. 

The satellite orbits and time epochs are calculated in the Geocentric Celestial Reference System (GCRS) thanks to the Standards of Fundamental Astronomy (SOFA) routines~\cite{SOFA:2016-05-03}. Then, we calculate the theoretical proper time of the onboard clock $\tauGR$ -- predicted by GR -- by integrating the coordinate time to proper time transformation:
\begin{equation}
	\tauGR = \int \frac{d\tau}{dt} dt = \int \left[ 1 - \frac{v^2}{2c^2} - \frac{U_E+U_T}{c^2} \right] dt
	\label{eq:transfo}
\end{equation}
where $\tau$ and $t$ are the proper time and the coordinate time (geocentric coordinate time TCG) of the clock, respectively, $c$ is the velocity of light in vacuum, $v$ is the velocity of the clock in the GCRS. Also, $U_E$ is the Newtonian gravitational potential of the Earth at the location of the satellite
\begin{equation}
	U_E = \frac{GM}{r} + \frac{GMR_0^2J_2}{2r^3} \left(1 - 3\cos^2\theta \right)
	\label{eq:pot}
\end{equation}
where $G$ is the gravitational constant, $M$, $R_0$ and $J_2$ are the mass, the equatorial radius and the zonal coefficient of order 2 of the Earth, respectively, and $r$ and $\theta$ are the distance from the center of the Earth and the co-latitude of the satellite, respectively. 
$U_T$ is the tidal potential due to external bodies~\cite{Wolf1995}
\begin{equation}
	U_T = \sum_{A} GM_A \left[ \frac{1}{|\bm r - \bm r_A|} - \frac{1}{|\bm r_A|} - \frac{\bm{r} \cdot \bm{r}_A}{|\bm r_A|^3} 
	\right]
	\label{eq:tides}
\end{equation}
where $M_A$ is the mass of external body $A$, and $\bm r$ and $\bm r_A$ are respectively the position vectors of the satellite and external body $A$ in the geocentric frame. We take into account Moon and Sun, while other bodies can be neglected.

\begin{figure}[t]
	\centering
	\includegraphics[width=\linewidth]{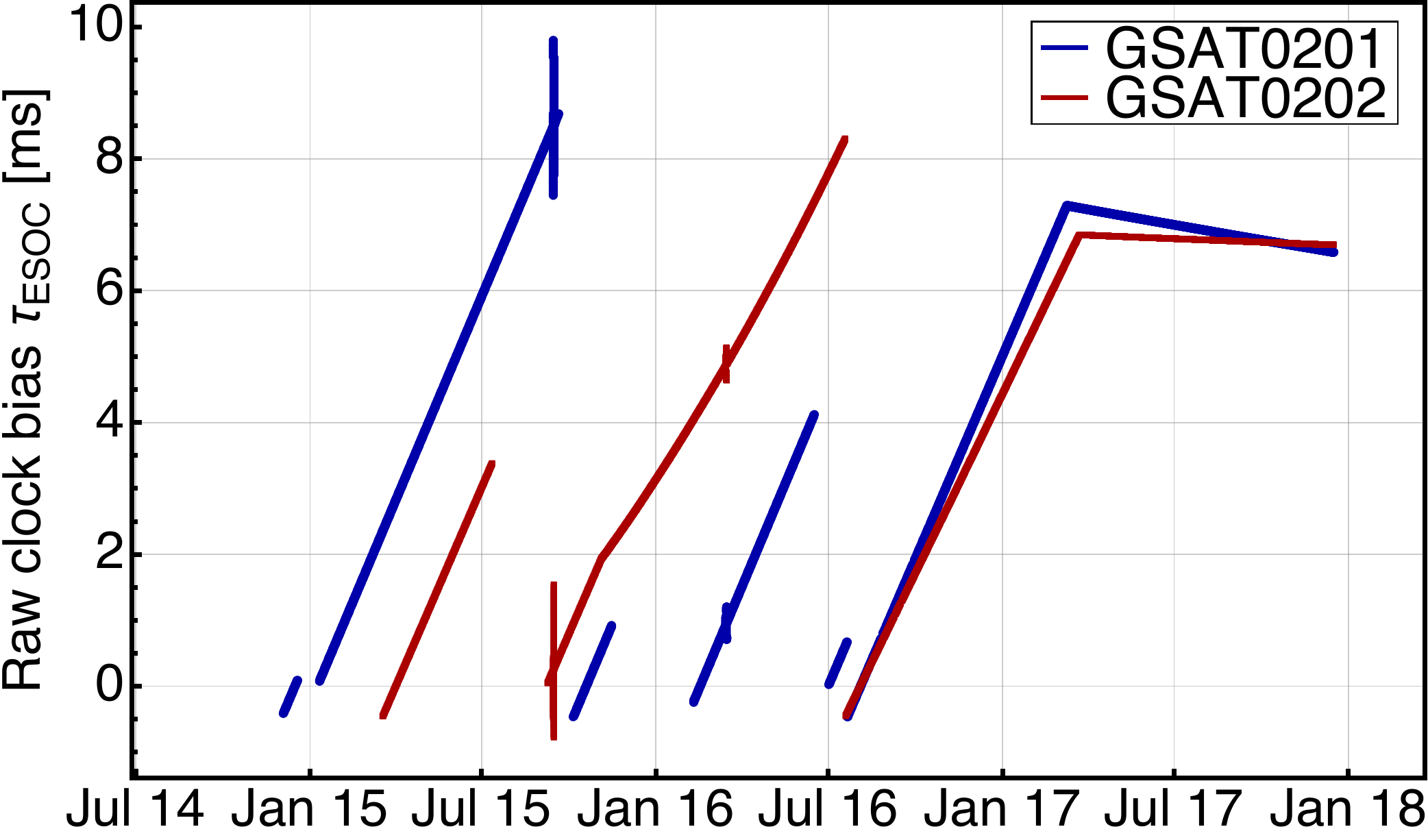}
	\caption{Raw clock bias $\tauData$, as read in the ESOC clock solution file.}
	\label{fig:clkbia}
\end{figure}

\begin{figure}[t]
	\centering
	\includegraphics[width=\linewidth]{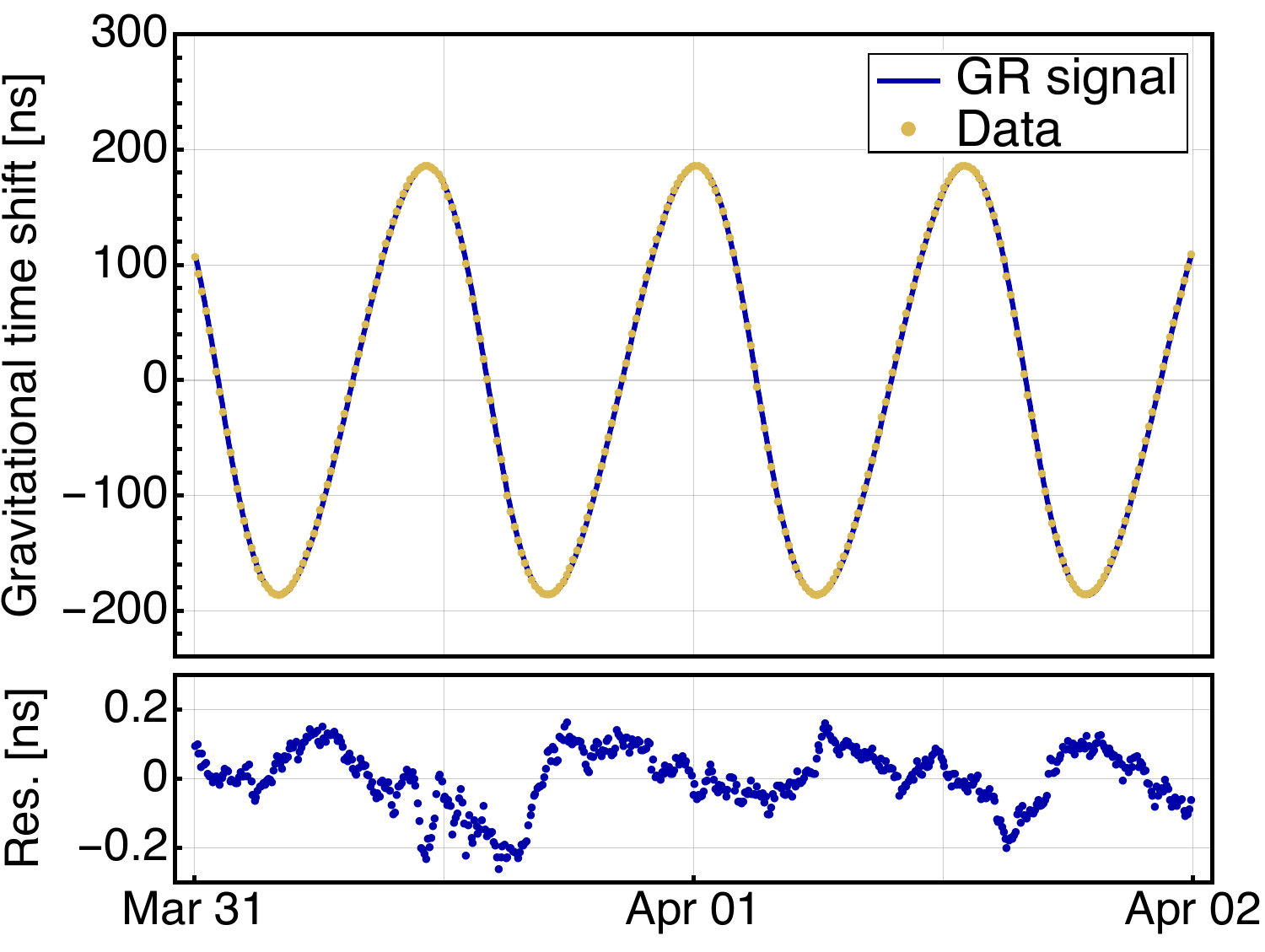}
	\caption{GR prediction, clock data (after removal of a daily linear fit) and residuals are shown for 2 days from March 31st, 2016. The peak-to-peak effect is around \SI{0.4}{\micro\second}, therefore the model and systematic effects at orbital period should be controlled down to \SI{4}{\pico\second} in order to have a \num{1e-5} uncertainty on the LPI violation parameter~$\alpha$.}
	\label{fig:timeshift}
\end{figure}

The main gravitational effect is the sum of a linear and a periodic term, which amounts to \SI{400}{\nano\second} peak-to-peak (see Figure~\ref{fig:timeshift}). The Earth flatness leads to a \SI{40}{\pico\second} peak-to-peak periodic effect at twice the orbital frequency. Tidal effects from the Moon and the Sun lead to a periodic signal of around \SI{12}{\pico\second} peak-to-peak, higher than the uncertainty goal of the experiment. 

The deviation of the proper time from the GR prediction, $\tauLPI$, is quantified by the LPI violation parameter $\alpha$ as given in (\ref{eq:redshift}) and proportional to the gravitational part of the coordinate to proper time transformation:
\begin{equation}
	\tauLPI = - \alpha \int \frac{U_E+U_T}{c^2} dt
	\label{eq:LPI}
\end{equation}

\begin{figure}[t]
	\centering
	\includegraphics[width=\linewidth]{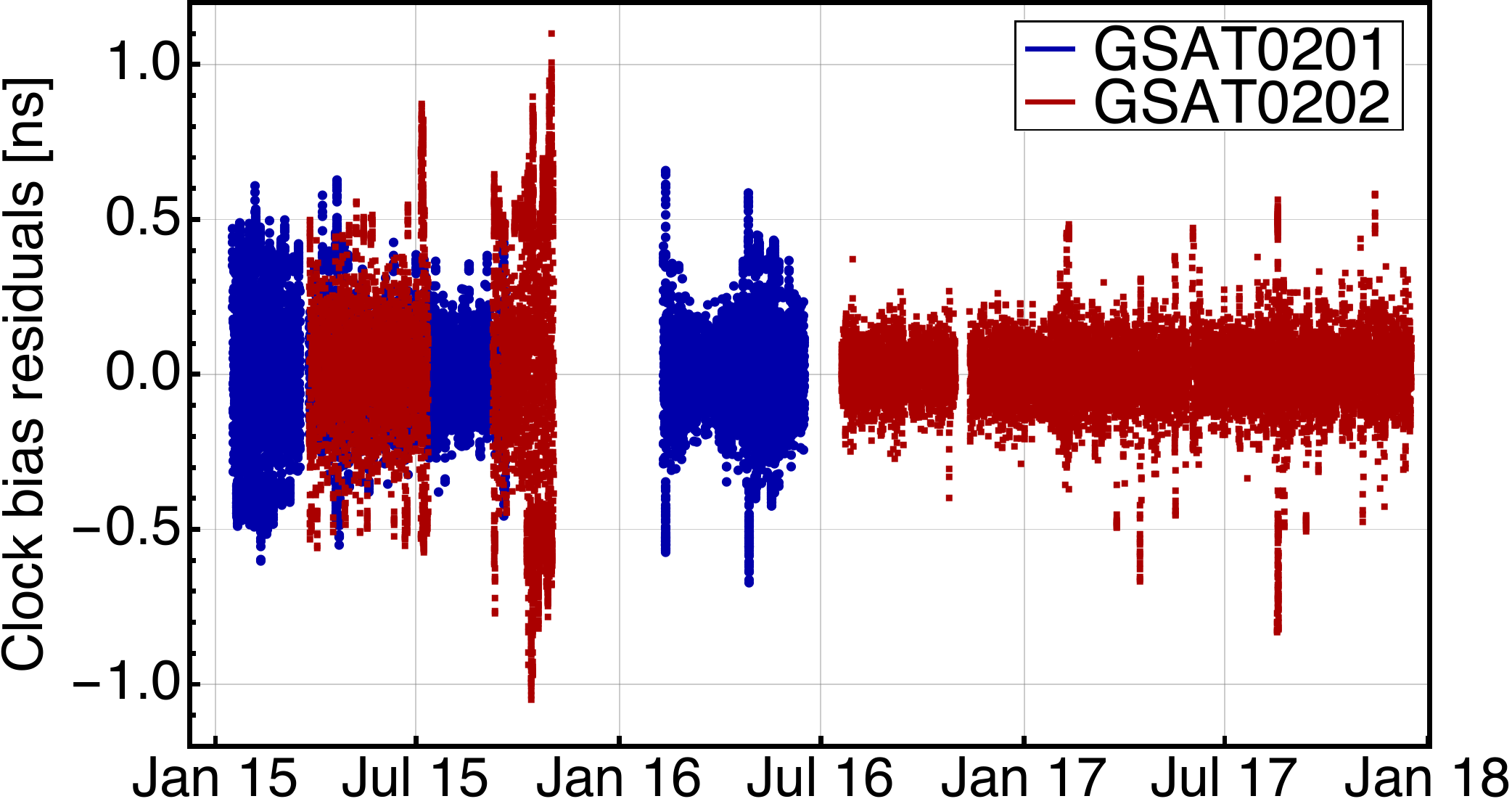}
	\caption{Clock bias pre-fit residuals are obtained by removing from the raw clock bias $\tauData$ a daily linear fit (DLF). Here only the times taken into account in the analysis are shown.}
	\label{fig:res}
\end{figure}

The raw clock bias $\tauData$ from the ESOC clock solution is shown on Figure~\ref{fig:clkbia} for satellites GSAT0201 and GSAT0202, with respect to a daily reference clock on the ground. A large drift of the order of \SI{34}{\micro\second\per\day} is present most of the time.
The linear part of the relativistic redshift between the Galileo clocks and a ground clock is $\approx \SI{40}{\micro\second\per\day}$ assuming a nominal 10.23 MHz frequency. However, each PHM clock is also affected by an intentional frequency offset ($\approx \SI{-6}{\micro\second\per\day}$) to this nominal frequency which explains the observed drift. Additionally, after each activation the clock retraces to the nominal frequency with an accuracy not better than $\pm \SI{0.18}{\micro\second\per\day}$. We account for this unknown frequency offset (together with the known $\approx \SI{34}{\micro\second\per\day}$) by removing from the clock bias a daily linear fit (DLF), which can be written in the form

\begin{equation}
	\tauDLF = \sum_{i=1}^{N} f_i(t) (a_i+b_i t)
	\label{eq:dlf}
\end{equation}
where $N$ is the number of days in the data, $a_i$ and $b_i$ are the clock offset and linear drift for day~$i$, respectively, and $f_i(t)$ is equal to~1 for day~$i$, and~0 otherwise. The clock bias residuals for the times chosen in the analysis are shown in Figure~\ref{fig:res}.

\begin{table}[t]
\centering
\caption{Master clock on board each eccentric satellite with dates and corresponding standard deviation of the clock bias pre-fit residuals. In bold are the chosen clocks for the gravitational redshift test.}
\label{table:clock}
\begin{tabular}{|l|l|c|c|c|}
\hline
Satellite                                      & clock & start & stop & \begin{tabular}[c]{@{}l@{}}clock residuals\\stand. dev. (ns)\end{tabular} \\ \hline
\multirow{2}{*}{GSAT0201}                       & {\bf PHM-B}\footnotemark[1] & 11/29/14 & 06/25/16 & 0.16 \\ \cline{2-5} 
                                               & PHM-A & 06/26/16 & 12/16/2017 & 0.69 \\ \hline
					       \multicolumn{1}{|c|}{\multirow{3}{*}{GSAT0202}} & {\bf PHM-B} & 03/17/15 & 11/03/15 & 0.20 \\ \cline{2-5} 
\multicolumn{1}{|c|}{}                         & RAFS  & 11/04/15 & 07/01/16 & 2.21 \\ \cline{2-5} 
\multicolumn{1}{|c|}{}                         & {\bf PHM-A}\footnotemark[2] & 07/02/16 & 12/16/2017 & 0.11 \\ \hline
\end{tabular}
\footnotetext[1]{GSAT0201 PHM-B was interrupted for 4 days as a master clock in favor of RAFS-B on 12/04/14. This data was removed.}
\footnotetext[2]{GSAT0202 PHM-A was interrupted for 13 days as a master clock in favor of RAFS-B on 10/30/16. This data was removed.}
\end{table}

The master clock on board the Galileo satellites may change over time due to maintenance routine. There are two PHM clocks as well as two rubidium clocks (RAFS) on board each of the satellites. In Table~\ref{table:clock} we show the dates of each master clock as well as the standard deviation of the corresponding clock residuals. We exclude from the analysis data from PHM-A of GSAT0201 and from RAFS of GSAT0202, because of the higher standard deviation of their residuals.
Obvious outliers at typically more than $10\sigma$ are removed, which represents around 7.7\% and 4.3\% of the total data for GSAT0201 and GSAT0202, respectively. Finally, our data analysis contains 359 days of data from GSAT0201 and 649 days of data from GSAT0202, spanning from January 2015 to December 2017. The raw clock bias $\tauData$ is corrected to account for the full GR prediction given in equation~(\ref{eq:transfo}), giving the corrected clock bias $\tauCorr$. This is explained in detail in the supplemental material~\cite{supp_mat}.

The data analysis is done in three steps. First, we fit a model for the stochastic noise to the corrected clock bias residuals. In a second step, we fit the model defined from equations~(\ref{eq:LPI}) and~(\ref{eq:dlf}) to the corrected clock bias by using a Monte Carlo approach, using the stochastic noise model estimated in the first step. This gives us the fitted value for $\alpha$ as well as an estimation of its statistical uncertainty. In a third step, we estimate the systematic uncertainty by considering the main sources of systematics: effects of magnetic field, of temperature and mismodelling of the orbital motion of the satellites.

The stochastic noise of the clock bias is modelled as a sum of white frequency and flicker phase noise (in the time domain that corresponds to power spectral density~(PSD) with $f^{-2}$ and $f^{-1}$ slopes, respectively), with the amplitudes given by a fit to the PSD of the clock bias residuals. The PSD is calculated thanks to the Lomb-Scargle algorithm~\cite{Lomb1976,Scargle1982}, which takes into account data gaps. Typical PSD noise levels are $\SI{3e-25}{\square\second\per\hertz}\times (f/f_0)^{-2}$ and $\SI{1e-21}{\square\second\per\hertz}\times (f/f_0)^{-1}$, where $f_0=\SI{1}{\hertz}$. As discussed in~\cite{Delva2015q} the clocks are also subject to flicker frequency noise at low frequencies (typically $\leq \SI{1}{\per\day}$), which (anyway) plays no role in our analysis because it is absorbed by the daily linear fit $\tauDLF$ given in (\ref{eq:dlf}).

As the noise from the clock bias is mostly composed by random walk noise, it is not possible to use a simple linear least-square approach which assumes white noise, and would lead to a strong under-estimation of the parameter uncertainties by one or more orders of magnitude. Therefore a Monte-Carlo linear least-square (MC-LLS) approach is used. The LLS minimizes the quantity $S(p) = (y-f(p))^\top (y-f(p))$, where $p$ is the set of parameters, $y$ is the observation vector, and $f(p)$ is the model estimated at $p$~(see e.g.~\cite[Chapter 15.6]{Press:2007:NRE:1403886}). In our case, $y=\tauCorr$, $f(p) = \tauLPI+\tauDLF$, and $p\equiv\{\alpha,a_i,b_i\}$, which are the same parameters as defined in equations~(\ref{eq:LPI}) and~(\ref{eq:dlf}). Moreover, the two clocks from GSAT0202 are weighted following their respective clock residuals standard deviations given in Table~\ref{table:clock}. This provides our estimates of the parameters $p$. Then we determine the statistical uncertainties of the parameters with the MC routine: we generate 1000 independent noise series mimicking our data, and fit the same model $f(p)$ to each of them, as to the data. This provides 1000~sets of the parameters $p$, coming only from the modelled stochastic noise. The standard deviation of the obtained parameter values give their statistical uncertainty at $1\sigma$.

We report the results of the MC-LLS, i.e.~the value of the LPI violation parameter $\alpha$ and its statistical uncertainty, in Table~\ref{tab:result}. We obtain $\alpha = \num{-0.77 \pm 1.48 e-5}$ and $\alpha = \num{6.75 \pm 1.41 e-5}$ for satellites GSAT0201 and GSAT0202, respectively. The value of $\alpha$ for GSAT0202 is~5 times its uncertainty at $1\sigma$, and therefore significant. A careful analysis of systematic effects is discussed in the supplemental material to explain this value~\cite{supp_mat}. We compared the MC-LLS approach to a General Least Square (GLS) approach for GSAT0201, where we take into account the full noise covariance matrix on a day by day basis. The value of $\alpha$ found with GLS is still consistent with a null value within the $1\sigma$ uncertainty, and the uncertainty found with GLS is 20\% smaller. However, the GLS neglects long term (across day boundaries) correlations and we consider the MC-LLS uncertainty value to be more conservative.

The main likely systematic effects were identified in~\cite{Delva2015q}. Effects acting on the frequency of the reference ground clock, as well as effects acting on the radio link can be safely neglected, as explained in~\cite{Delva2015q}. We will assess effects acting directly on the frequency of the onboard clock, namely temperature and magnetic field variations, as well as systematic effects coming from orbit modelling errors, which are strongly correlated to the clock solution in the case of a one-way time transfer (see e.g.~\cite{Montenbruck2014,Svehla2014}). During this experiment, no additional environmental data (onboard magnetic field or temperature) was available.
Therefore, we will only evaluate an upper limit of the systematic effects rather than trying to correct them. In doing so we do not use the clock data itself, so our limits are independent of a putative violation of the gravitational redshift.

A detailed description of the systematic effect analysis is given in the supplemental material~\cite{supp_mat}. Here we summarize the main results. The magnetic field vector is calculated along the trajectory of each satellite, and projected onto each axis of the PHM clock. The sensitivity of the clock to the magnetic field, as determined in ground tests, then translates the modelled magnetic field variations along each axis of the clock into a variation of the fractional frequency of the clock. The model $f(p) = \tauLPI+\tauDLF$ is then fitted to this variation to obtain the highest possible value of $\alpha$ due to this effect. Our approach is conservative as we do not assume any shielding from the satellite or the clock. The result is reported in Table~\ref{tab:result} in the magnetic field uncertainty column. 
A similar approach is used to estimate the highest possible value of $\alpha$ due to temperature variations, acting both directly on the clock and on the rest of the payload. We assume that temperature variations of the clock are due to the change of the orientation of the satellite w.r.t.~the Sun, and take their amplitude as the highest peak-to-peak variation allowed by the thermal control system, which is a very conservative assumption as explained in the supplemental material~\cite{supp_mat}. The result is reported in Table~\ref{tab:result} in the temperature uncertainty column. 
Finally, we estimate uncertainties due to orbit modelling errors thanks to satellite laser ranging (SLR) data. Indeed, SLR residuals have been shown to be highly correlated to the clock bias, as it is expected in a one-way time transfer~\cite{Montenbruck2014,Svehla2014}. We fit the same model $f(p) = \tauLPI+\tauDLF$ to the (scaled) SLR residuals in order to get the highest value of $\alpha$ due to orbit modelling errors. The result is reported in Table~\ref{tab:result} in the orbit uncertainty column.

When we quadratically add the statistical and systematic uncertainties due to each considered error source, we obtain for the LPI violation parameter $\alpha = \num{-0.77 \pm 2.73 e-5}$ for GSAT0201 and $\alpha = \num{6.75 \pm 5.62 e-5}$ for GSAT0202 (see Table~\ref{tab:result}). 

Finally, we combine the data from both satellites using a global MC-LLS analysis, where the only parameter common to both satellites is the LPI violation parameter $\alpha$. 
The relative weight of both satellites in the MC-LLS is chosen following their orbit uncertainty in Table~\ref{tab:result}.
The uncertainties coming from systematics are evaluated in the same way as for each satellite alone, except that we combine the modelled clock variations due to systematics from both satellites in the fit, with the same weight as for the clock biases. The results are reported in Table~\ref{tab:result}.

To conclude, by analysing 1008~days of data from the two eccentric Galileo satellites, GSAT0201 and GSAT0202, and through a careful analysis of systematic effects, we were able to improve the gravitional redshift test done by GP-A in 1976 by a factor 5.6, down to $\alpha = \num{+0.19 \pm 2.48 e-5}$. Our result is at the lower edge of the predicted sensitivity in~\cite{Delva2015q}. This is due to the very favourable configuration of GSAT0201 with respect to the orbit systematics on the clock bias, which is almost \ang{90} out-of-phase with the LPI violation signal. At this point, the main residual limiting factor is the uncertainty due to the magnetic field variations, which cannot be overcome without more information about the clock sensitivity (e.g.~directional dependence) and the actual local magnetic field after e.g.~shielding from the satellite itself. A refinement of the magnetic field characterisation of the PHM per axis could be performed to improve the magnetic field contribution uncertainty and reduce further the LPI overall total uncertainty. In any case, we can see that the three main uncertainties, i.e., statistical, orbit and magnetic field, are of the same order. Therefore, envisaging a potential future mission of the same type, it would be of interest to improve these three aspects of the experiment: a more stable clock to have better statistics, a careful shielding, modelling or measurement of the magnetic field, and a careful modelling or measurement of non-gravitational accelerations. Also increasing the signal (higher ellipticity, lower perigee) would improve the test significantly (see e.g.~the STE-QUEST proposal~\cite{Altschul2015a}). Finally, a two-way link would strongly reduce the effect of orbit determination uncertainties (see e.g.~the ACES proposal~\cite{Cacciapuoti2009a,Meynadier2018}).

\onecolumngrid

\begin{table}[h] 
	\centering
	\begin{tabular}[t]{|c|c|c||c|c|c|c|}
		\hline
		& {\bf LPI violation} & {\bf Total uncertainty} & Stat. unc. & Orbit unc. & Temperature unc. & Magnetic field tot. unc. (X/Y/Z)\\
		& {\bf [$\times10^{-5}$]} & {\bf [$\times10^{-5}$]} & [$\times10^{-5}$]  & [$\times10^{-5}$] & [$\times10^{-5}$] & [$\times10^{-5}$] \\\hline\hline
		\rule[-0.2cm]{0cm}{0.6cm} GSAT0201 & $-0.77$ & $2.73$ & $1.48$ & $1.09$ & $0.59$ & $1.93$ \ $(0.52/-0.36/1.82)$  \\\hline
		\rule[-0.2cm]{0cm}{0.6cm} GSAT0202 & $6.75$ & $5.62$ & $1.41$ & $5.09$ & $0.13$ & $1.92$ \ $(-0.07/0.58/1.83)$ \\\hline\hline
		\rule[-0.2cm]{0cm}{0.6cm} \bf{Combined} & $\bm{0.19}$ & $\bm{2.48}$ & $1.32$ & $0.70$ & $0.55$ & $1.91$ \ $(0.48 / -0.29 / 1.82)$ \\\hline
	\end{tabular}
	\caption{Final result of the EEP/LPI test. Each row is a separate output from fits of respectively GSAT0201 data and models, GSAT0202 data and models, and a joint fit of both sets of data and models. The uncertainties due to systematic effects are evaluated independently from the clock data, and are the result of individual fits of models of systematics (for temperature and magnetic field), and a fit of SLR residuals for the orbit uncertainty, thus giving an upper limit of each effect. A single value is computed for the magnetic field uncertainty by summing in quadrature the X/Y/Z values. The total uncertainty column, for each row, is derived from the quadratic sum of the individual uncertainties to the right.}
	\label{tab:result}
\end{table}

\twocolumngrid

\begin{acknowledgments}
	The authors would like to thank S. Loyer (CNES/CLS) and Krzysztof So\'snica (Wroc{\l}aw University of Environmental and Life Sciences) for their precious help. We thank C.~E. Noll from the ILRS Central Bureau for organizing the GREAT SLR campaign, and all ILRS stations who participated, and in particular Grasse, Herstmonceux, Wettzell and Yarragadee. We thank people from ESA for their help at many levels: T. Springer and F. Gini from ESA/ESOC; L. Cacciapuoti, M. Armano, D. Navarro-Reyes, J. Hahn and P. Waller from ESA/ESTEC; M. Castillo from ESA/ESAC; I. Carnelli and C. de Matos from ESA/HQ; A. Bauch, M. Rothacher and F. Vespe for their support as members of the ESA GNSS Science Advisory Committee (GSAC). We acknowledge financial support from ESA within the GREAT (Galileo gravitational Redshift Experiment with eccentric sATellites) project, from Paris Observatory/GPhys specific action and from Generalitat Valenciana APOSTD2017. Part of the data used in this article will be published by ESA on the ESA GNSS Science Server at \url{https://gssc.esa.int/} during 2018.
\end{acknowledgments}


%

\end{document}